\documentclass[aps,prl,preprint,groupedaddress]{revtex4-2}
\usepackage{graphicx}
\usepackage{newtxtext}
\usepackage{newtxmath}
\usepackage{hyperref}
\usepackage{bm}
\usepackage{xcolor}
\usepackage{lineno}
\hypersetup{
    colorlinks = true,
    colorlinks=true,
    linkcolor=blue,
    filecolor=blue,
    urlcolor=blue,
    citecolor=blue,
}

\newcommand{\RomanNumeralCaps}[1]

\begin{document}
\setcounter{secnumdepth}{3}

\title{Rising bubbles draw surface patterns: a numerical study}


 
\author{Dabao Li}
\affiliation{School of Engineering Science, University of Chinese Academy of Sciences, Beijing 101408, China}
\author{Lang Qin}
\affiliation{State Key Laboratory of Hydroscience and Engineering, and Department of Energy and Power Engineering, Tsinghua University, Beijing 100084, China}
\author{Zhigang Zuo}
\affiliation{State Key Laboratory of Hydroscience and Engineering, and Department of Energy and Power Engineering, Tsinghua University, Beijing 100084, China}
\author{Guangzhao Zhou}
\email{zgz@ucas.ac.cn}
\affiliation{School of Engineering Science, University of Chinese Academy of Sciences, Beijing 101408, China,}
\affiliation{State Key Laboratory of Nonlinear Mechanics, IMECH \& UCAS, Beijing 100190, China}


\date{\today}

\begin{abstract}

Small bubbles rising in a chain can self-organize into regular patterns upon reaching a liquid's free surface.
This phenomenon is investigated through direct numerical simulations.
By varying the bubble release period, distinct branching patterns characterized by different numbers of arms are observed.
These macroscopic patterns are found to stem from localized interactions between the bubbles, the free surface, and the surrounding liquid flow.
A heuristic model is proposed that quantitatively relates the number of arms to the bubble release period and predicts the probability of observing specific arm counts.
This study provides insights into broader nonlinear pattern formation and self-organization phenomena.

\end{abstract}


\maketitle

\section{Introduction}
\label{sec:intro}

The dynamics of gas bubbles rising in quiescent liquids have long been a fundamental topic in gas-liquid two-phase flow research\citep{Amol2005, Takagi2011, Lohse2013, Lohse2018, Risso2018}.
Extensive investigations have been conducted on various aspects of single bubble behavior during ascent, including bubble morphology and terminal velocity\citep{Raymond2000, Takagi2003, Tripathi2015}, wake flow characteristics\citep{Vries2002}, and trajectory instability\citep{Mougin2001, Shew2006, Risso2012}.
In practice, a probably more prevalent phenomenon is the formation of bubble chains, as commonly observed in a glass of champagne or sparkling water.
In such systems, millimeter-scale bubbles are periodically generated at discrete nucleation sites located at the bottom wall of the vessels.
Bubbles within a chain exhibit distinct dynamics compared to individual rising bubbles, such as a higher terminal velocity primarily due to the influence of wake flows generated by preceding bubbles\citep{Marks1973, Zhang2003, Wang2015}.
The wake flow can also induce unsteady lift forces, which may destabilize the configuration of the bubble chain\citep{Atasi2023}.

When a bubble ascends to the gas-liquid interface, it may rapidly burst, accompanied by generation of small droplets\citep{Lhuissier2012,Ghabache2016,Deike2018,Gordillo2019}.
In cases where liquid viscosity is large or when contaminants/surfactants are present, bubbles can float and persist for a considerable time\citep{Sanada2005,Poulain2018}.
The complexity of bubble-free surface interactions becomes more pronounced in systems that are specially designed and controlled\citep{Feng2016,Guan2025}.

\begin{figure}
	\centering
    \includegraphics[width=0.8\linewidth]{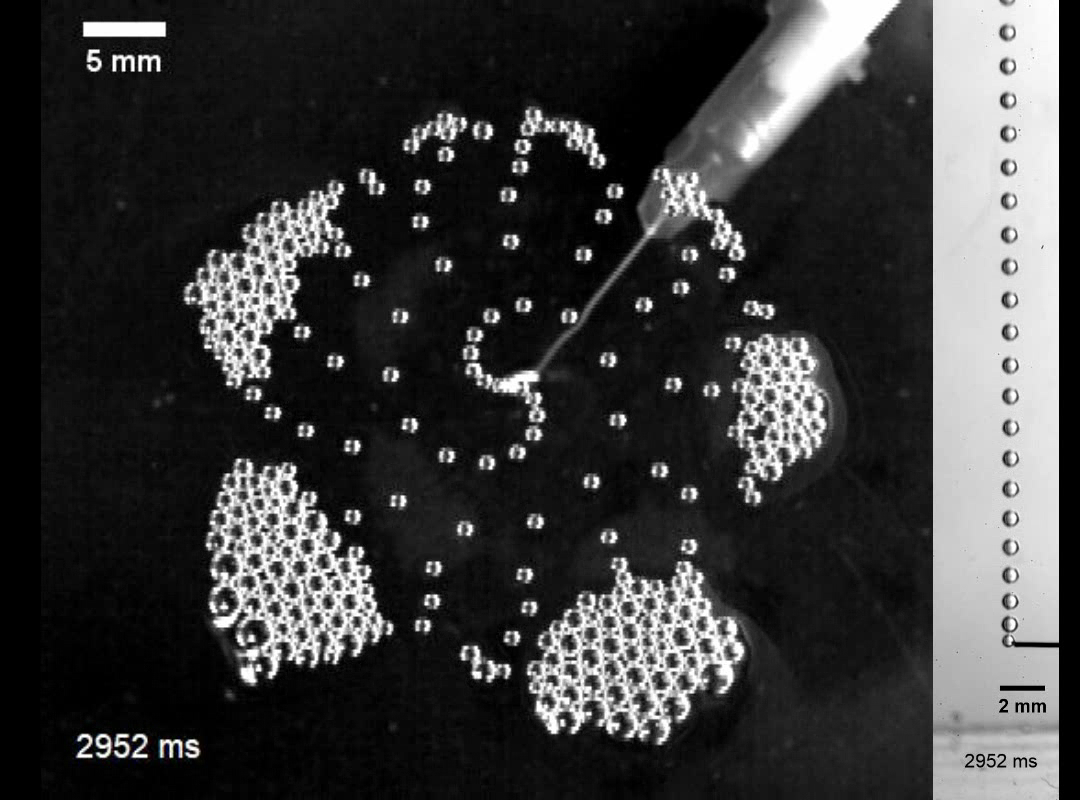}
    \caption{Top view of the surface pattern formed by periodically injected air bubbles into a vessel partially filled with silicone oil.
    The image on the right is the side view of the bubble chain at the same instant.}
	\label{fig:experiment}
\end{figure}

This study centers on the collective behavior of periodically-released small bubbles rising in a chain toward the free surface.
After reaching the interface, the bubbles exhibit lateral migration within the horizontal plane, forming spatial patterns observed from above.
While most of these patterns appear disordered, under specific conditions, the bubbles spontaneously self-organize into stable, regular configurations with rotational symmetry around the vertical ascent trajectory.
An experimental demonstration of these regular surface patterns is shown in Fig.~\ref{fig:experiment}.
In this experiment, air bubbles with a uniform diameter of approximately $0.8\,\mathrm{mm}$ are periodically injected from a syringe into a vessel partially filled with silicone oil.
A spiral pattern is clearly observed on the free surface when looking from above.
The density, dynamic viscosity, and surface tension coefficient of the silicone oil are $935\,\mathrm{kg/m^3}$, $0.01\,\mathrm{Pa{\cdot}s}$, and $0.02\,\mathrm{N/m}$, respectively.
The bubble release interval is $16\,\mathrm{ms}$.

Despite the likelihood that such phenomenon has been observed since very early times, the literature on it is limited.
The first article on this specific topic appears to be \citet{Yoshikawa2010}, where the authors conducted an experimental investigation of the surface bubble patterns as an analogy to the regular leaf arrangement in the context of phyllotaxis study (see, e.g., Ref.~\citep{Douady1992}).
A phenomenological model was later developed by the same research group\citep{Yoshikawa2012}.
These two papers acknowledge the importance of bubble-bubble interactions and flow advection.
Nevertheless, the way in which these factors contribute to pattern formation require further elucidation.
In this paper, we revisit these regular surface patterns ``drawn'' by rising bubbles from a numerical perspective, aiming to gain a deeper understanding of the essential underlying physics of this self-organized phenomenon.

The rest of the paper is arranged as follows: Section~\ref{sec:numerical} describes the numerical setup of the simulations.
This is followed by section~\ref{sec:result}, where the numerical results are presented and discussed.
We further propose a heuristic model in section~\ref{sec:theory}.
A summary is given in section~\ref{sec:conclusion}.

\section{Numerical setup} 
\label{sec:numerical}

We use the open-source software Aphros\citep{aphros2022} for direct numerical simulations of the present two-phase flow system.
Aphros solves the time-dependent incompressible Navier-Stokes equations with the divergence-free constraint, which read
\begin{equation}
\begin{aligned}
\rho\left[\frac{\partial \bm{u}}{\partial t}+(\bm{u} \cdot \nabla) \bm{u}\right] & =-\nabla p+\nabla \cdot \left[ \mu  \left(\nabla \bm{u}+\nabla \bm{u}^T\right) \right] +\bm{f}_\sigma+\rho \bm{g},
\label{eq:NS2}
\end{aligned}
\end{equation}
and
\begin{equation}
\begin{aligned}
\nabla \cdot \bm{u} & =0,
\label{eq:NS1}
\end{aligned}
\end{equation}
respectively, where $t$ is time, $\bm{u}$ is the velocity vector, $p$ is pressure, $\bm{g}$ is the gravitational acceleration whose magnitude is denoted by $g$, and $\bm{f}_\sigma$ represents the surface tension effect.
$\rho$ and $\mu$ are the effective density and dynamic viscosity, respectively.
They are calculated as weighted averages based on the local gas volume fraction $\alpha$: 
\begin{equation}
\rho=(1-\alpha) \rho_l+\alpha \rho_g,
\quad \mu=(1-\alpha) \mu_l+\alpha \mu_g,
\end{equation}
where $\rho_l$, $\rho_g$, $\mu_l$, and $\mu_g$ represent the density and viscosity of the liquid and gas phases, respectively.
The evolution of $\alpha$ is governed by the advection equation:
\begin{equation}
\frac{\partial \alpha}{\partial t}+(\bm{u} \cdot \nabla) \alpha=0.
\end{equation}
Eqs.~\eqref{eq:NS2} and \eqref{eq:NS1} are solved using the second-order Bell-Colella-Glaz method\citep{Bell1989}.
The volume fraction field advection is handled using the volume-of-fluid (VOF) method, with a piecewise linear interface calculation (PLIC) method\citep{Scardovelli2000} employed for interface reconstruction.
The surface tension term is defined as $\bm{f}_\sigma=\sigma \kappa \nabla \alpha$, where $\sigma$ denotes the surface tension coefficient.
The interfacial curvature, $\kappa$, is evaluated using the method of particles\citep{Karnakov2020}.

Additionally, an improved multi-marker technique\citep{aphros2022} is employed in the VOF method to prevent bubble coalescence, ensuring that bubbles can float on the free surface without bursting in the simulation.
It is noteworthy that the present study focuses on cases in which bubble coalescence and bursting are absent.
Under these conditions, the multi-marker method
ensures fidelity in capturing bubble-bubble and bubble-free surface interactions without introducing non-physical
artifacts\citep{Coyajee2009,aphros2022}.
The simulation results and conclusions would not be valid if coalescence or bursting were to occur.

The accuracy of the solver is validated by simulating a single bubble rising in quiescent liquid.
Four different liquids are tested with their physical parameters the same as those of cases S3, S5, S6, and S8 in Ref.~\citep{Raymond2000}.
The Morton numbers, defined as $\mathit{Mo}=g\mu_l^4/(\rho_l\sigma^3)$, are  0.11, $9\times 10^{-4}$, $1\times 10^{-4}$, and $9\times 10^{-7}$ for these four cases, respectively.
For each liquid, the volume-equivalent spherical diameter of the bubble, $d$, is varied across a relatively wide range.
The computational domain is a liquid-filled box with a Cartesian mesh composed of cubic cells.
The cell size is set to approximately $d/10$, maintaining consistent spatial resolution for different bubble sizes.

\begin{figure}
	\centering
    \includegraphics[width=.5\linewidth]{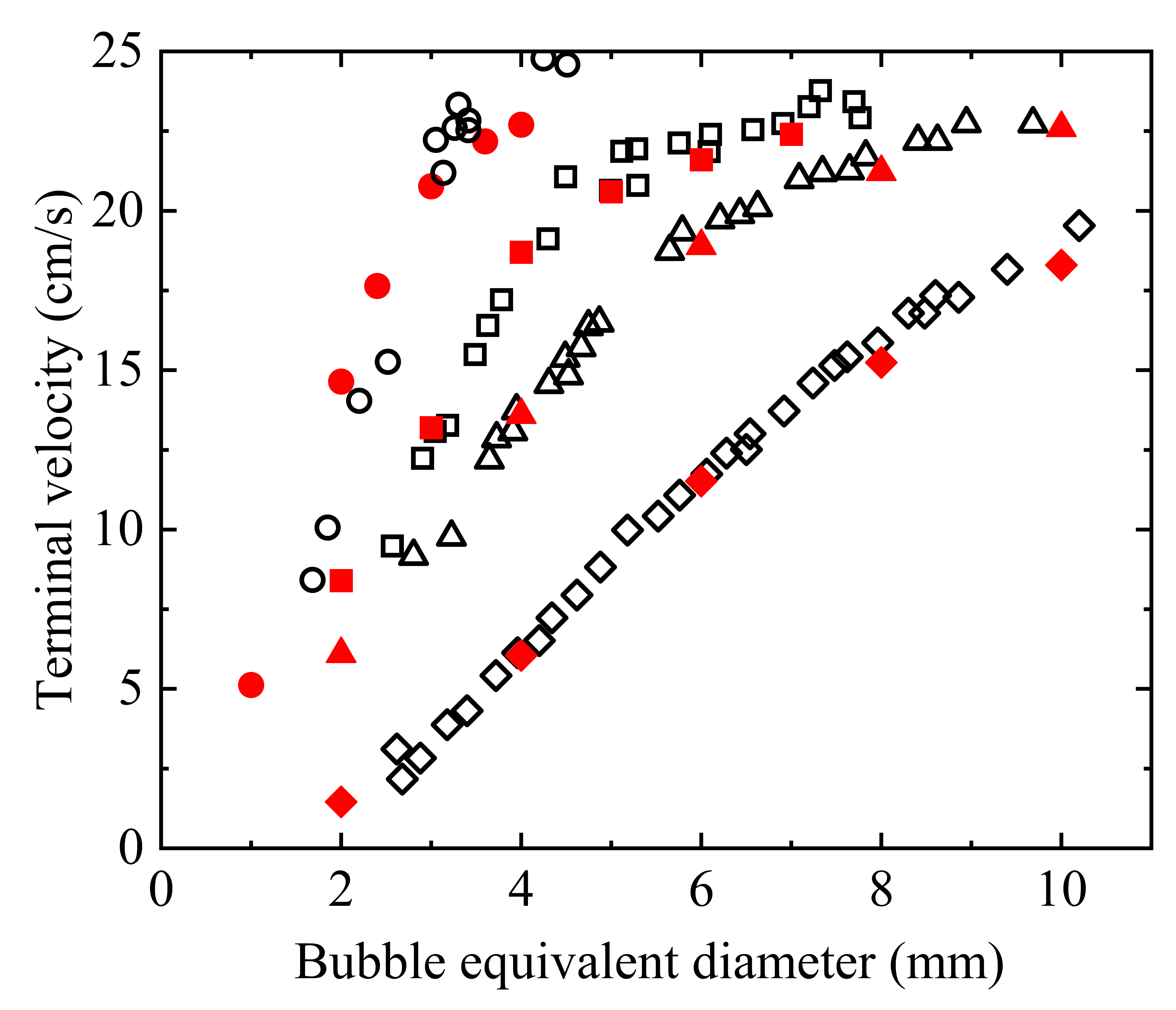}
    \caption{Bubble's terminal rise velocity versus equivalent diameter.
    The black hollow symbols represent the experimental data sets S3 (diamonds), S5 (triangles), S6 (squares), and S8 (circles) from Ref.~\citep{Raymond2000}.
    The Morton numbers for these four cases are $0.11$, $9\times10^{-4}$, $1\times10^{-4}$, and $9\times10^{-7}$, respectively.
    The red solid symbols with the same shapes are the corresponding simulation results using Aphros.
    }
	\label{fig:verify1}
\end{figure}

As can be seen in Fig.~\ref{fig:verify1}, the simulated terminal velocities of the bubbles (red solid symbols) show good agreement with experimental data from Ref.~\citep{Raymond2000} (black hollow symbols).
Notably, the terminal velocity is influenced by the deviation of the bubble shape from a sphere.
Therefore, the results in Fig.~\ref{fig:verify1} also demonstrate the solver's capability in capturing bubble deformation during ascent.
More validation cases can be found in Ref.~\citep{aphros2022}.

\begin{figure}
	\centering
    \includegraphics[width=.85\linewidth]{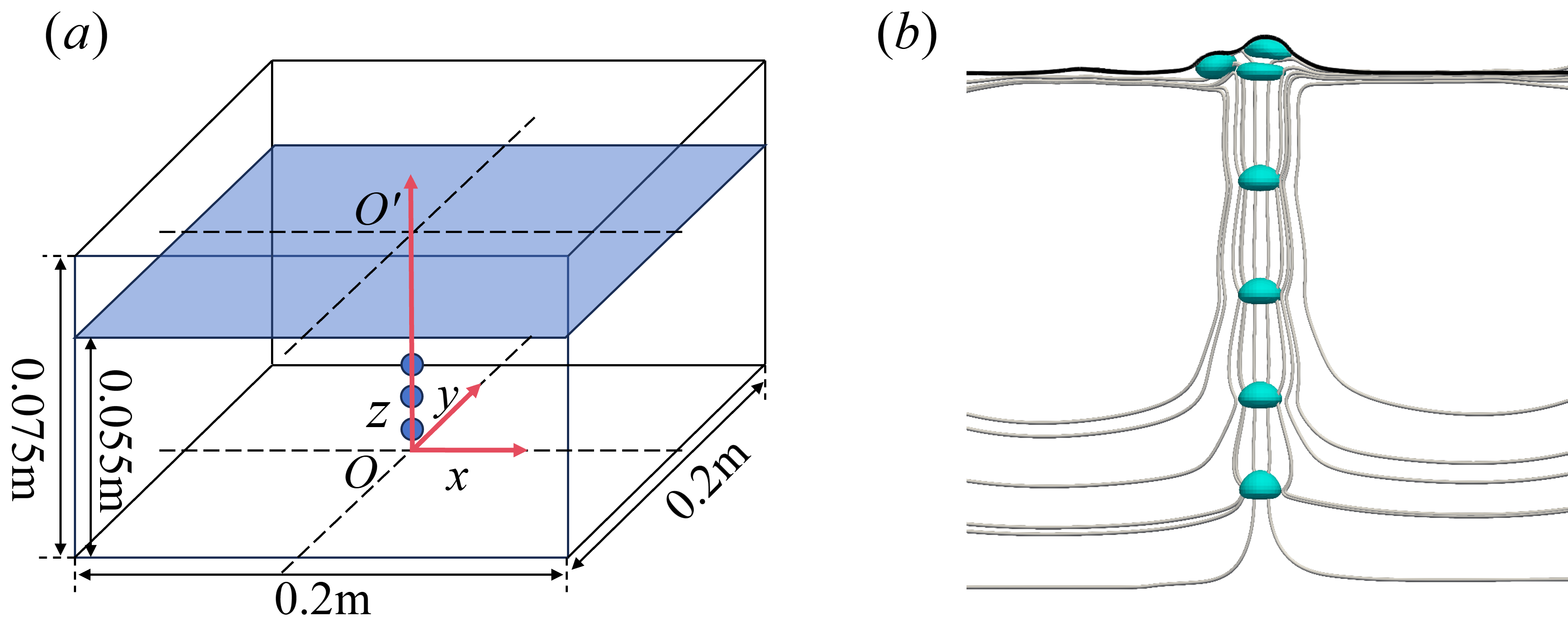}
    \caption{(a) Illustration of the computational domain (bubble size not to scale) and the coordinate system.
    (b) Simulated bubbles rising in a chain, with the free surface (black) and streamlines (gray) shown in a cross section through the $z$-axis. Only a small region close to the bubble chain is displayed.}
	\label{fig:overview}
\end{figure}

The computational domain for simulations of the surface patterns is a square box partially filled with liquid, see Fig.~\ref{fig:overview}(a).
A Cartesian coordinate system is established with the origin $O$ located at the geometric center of the box's bottom wall. 
The $z$-axis is oriented upward, while the $x$- and $y$-axes align with the side walls of the box.
The dimensions of the box are $[-0.1,0.1]\,\text{m}\times[-0.1,0.1]\,\text{m}\times [0,0.075]\,\text{m}$, 
with the undisturbed free surface located at $z=0.055\,$m.
Gas bubbles with a volume-equivalent diameter of $d=4$\,mm are released periodically from a source at $(0,\, 0, \, 0.005)$\,m.
We generate the bubbles by setting the gas volume fraction to 1 in a spherical region centered at the source.
A uniform computational mesh with $N_x \times N_y \times N_z = 512 \times 512 \times 192$ cubic cells is used, ensuring that there are about 10 mesh cells within the diameter of a spherical bubble.
At this resolution, each case running in parallel on 128 CPU cores requires approximately four days.
A finer mesh is also tested, yielding consistent results.
More details can be found in Appendix~\ref{sec:appendix}.

Free-slip conditions are imposed at the top and bottom boundaries of the computational domain, while periodic conditions are applied on all side boundaries.
In our simulations, the side boundaries are sufficiently far from the area of interest, namely, the central region of the free surface.
Therefore, their influence on the bubble dynamics is negligible.
It should be noted, however, that when side walls are close to the bubble chain, the vertical bubble motion can induce significant global recirculation in the confined space.
This confined recirculation may subsequently alter the bubble velocity and, consequently, the final pattern formation.
For the single-bubble cases shown in Fig.~\ref{fig:verify1}, a bubble acquires terminal velocity soon after release.
However, because of the wake effects from preceding bubbles, for a bubble rising in a bubble chain, a much larger height is required for its velocity to reach a constant value.
In the simulations of the current study, the bubbles are still gradually accelerating before approaching the free surface.
Consequently, the liquid depth $h$ (or its dimensionless form $h/d$) is also a relevant parameter.
Its effect will be briefly discussed in section~\ref{sec:conclusion}.

In all simulations, the gravitational acceleration is set to $g=10\,\mathrm{m/s^2}$ in the $-z$ direction.
The density and dynamic viscosity of the liquid ($l$) and gas ($g$) phases are: $\rho_l = 1000\,\mathrm{kg/m^3}$, $\rho_g = 30\,\mathrm{kg/m^3}$, $\mu_l=2\times10^{-2}\,\mathrm{Pa\cdot s}$, and $\mu_g=2\times10^{-5}\,\mathrm{Pa\cdot s}$.
The surface tension coefficient $\sigma=0.02\,\mathrm{N/m}$.
These physical parameters are similar to those used by \citet{Yoshikawa2010}, allowing a reasonable comparison between the numerical results and the experimental data in Ref.~\citep{Yoshikawa2010}.  
Given the low density ratio ($\rho_g/\rho_l=0.03$) and viscosity ratio ($\mu_g/\mu_l=0.001$) between gas and liquid, the effects of gas phase properties are expected to be minimal.
The Ohnesorge and Bond numbers are calculated as $\mathit{Oh} = \mu_l/\sqrt{\rho_l \sigma d} \approx 0.071$ and $\mathit{Bo} =\rho_{l}gd^2/\sigma=8$, respectively.
The corresponding Morton number is $\mathit{Mo}=\mathit{Oh}^4 \mathit{Bo}=2\times 10^{-4}$, lying between the values of cases S5 and S6 in Fig.~\ref{fig:verify1}.
The bubbles exhibit moderate deformation during ascent.
No path instabilities are observed with the parameters investigated.

\section{Results and discussion} \label{sec:result}

\subsection{Overview} \label{subsec:overview}

With the VOF method, the motion and deformation of each bubble are fully resolved.
Figure~\ref{fig:overview}(b) shows a cross-sectional view of the numerical result for a typical case we have simulated.
The bubbles rise along a straight path in a chain and are arrested by the free surface. 
As can be seen from the streamlines, the bubble chain induces a vertical liquid jet along the central axis.
This jet later on converts into a divergent radial flow in a thin layer beneath the free surface.
Owing to the overshooting effect of the bubbles and jet, a small circular bump forms on the free surface, centered at the bubble emergence site.
These observations agree well with those reported in Ref.~\citep{Yoshikawa2012}.

\begin{figure}
	\centering
    \includegraphics[width=1\linewidth]{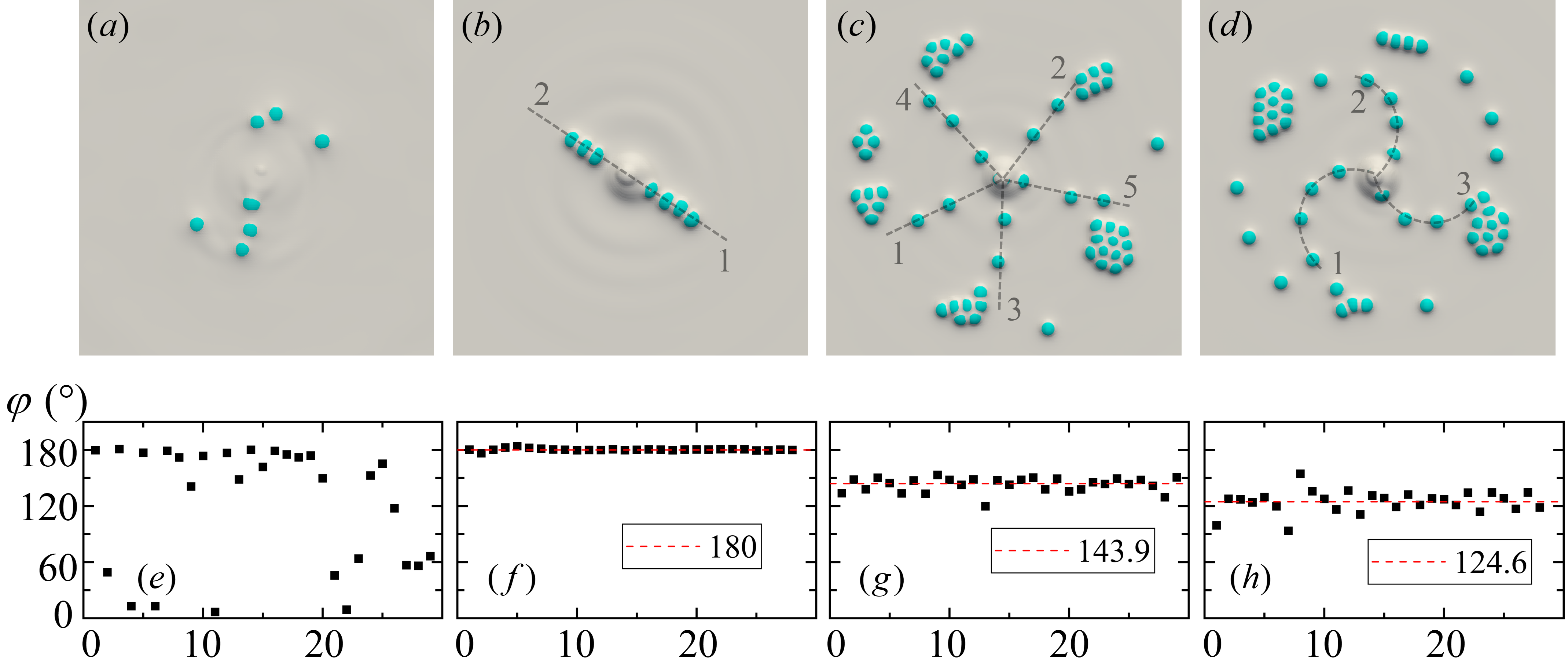}
    \caption{(a) to (d): Four typical surface patterns observed in the numerical simulations (top view). The free surface is colored in gray.
    The dashed lines serve as visual guides to distinguish between different branches.
    The numbers indicate the sequence in which successive bubbles join these branches.
    From left to right, the release periods are 100, 50, 42, and 36\,ms, respectively.
    (e) to (h): Divergence angle as a function of bubble release sequence number for the corresponding cases in the top row.
    The red dashed lines show the averaged values.
    }
	\label{fig:patterns}
\end{figure}

Systematic simulations are performed to examine the effect of the bubble release period $T$ on pattern formation, with other parameters held constant.
The top panels of Fig.~\ref{fig:patterns} depict four representative surface patterns observed in the simulations, with $T$ decreasing from left to right.
Panel (a) shows a disordered mode, characterized by a random distribution of bubbles on the free surface.
Panel (b) displays a two-armed mode, where the bubbles align along two branches in a straight line.
At smaller bubble release periods, multi-armed modes can be observed, as shown in panels (c) and (d).
The corresponding movies for all these four cases can be found in the Supplemental Material\citep{SM}.

It is known that the pattern formation is closely related to the directions in which the bubbles are emitted in sequence.
In the bottom panels of Fig.~\ref{fig:patterns}, the difference $\varphi$ 
between  the emission directions of consecutive bubbles, referred to as the ``divergence angle'' in Ref.~\citep{Yoshikawa2010}, is shown as a function of bubble release sequence number.
As will be subsequently elucidated, $\varphi$ is fully determined by the local dynamics in the vicinity of the bubble emergence site.
In detail, four major processes are identified to be responsible for the formation of different surface patterns.
They are (i) pair collisions, (ii) non-contact repulsion, (iii) bump confinement, and (iv) flow advection.
The first two processes both describe the interaction between bubbles.
Pair collision occurs when two bubbles are close enough with their mass center distance $\ell \lesssim d$, and is associated with significant bubble squeezing, deformation, and quick velocity changes.
Whereas non-contact repulsion is effective when $\ell \gtrsim d$ and is less violent.

For the sake of clarity, in the following, the specific bubble that is focused on is named bubble 0.
Other bubbles are labeled in chronological order of their release.
Specifically, the two bubbles that rise to the free surface prior to bubble 0 are bubbles $-2$ and $-1$, respectively.
The one that follows it is called bubble 1.

\subsection{Disordered and fixed two-armed modes}\label{subsec:disorderAndTwoArmed}
The disordered mode is seemingly the most common situation.
It occurs when the bubble release period $T$ is relatively large ($T \gtrsim 100\,\mathrm{ms}$ in the case illustrated in Fig.~\ref{fig:patterns}).
This mode is characterized by random pair collisions between bubbles.
For instance, as bubble $-1$ rises to the free surface, it collides with bubble $-2$ which is already at the emergence site.
The two bubbles are knocked away by each other in opposite directions, creating a $180^\circ$ divergence angle.
By the time bubble 0 reaches the surface, the influence of the previous collision has vanished.
Bubble 0 stays at the center and collides with the subsequent bubble 1, resulting in opposite emission directions that do not necessarily align with those of bubbles $-2$ and $-1$.
As a consequence, the divergence angle jumps between $180^\circ$ and other random values (Fig.~\ref{fig:patterns}e).
Additionally, a large $T$ results in a relatively weak liquid jet, such that the bubbles tend to aggregate near the central region of the free surface.
Both of these features hinder the formation of regular patterns with distinguishable numbers of arms.

The reason for the transition from the disordered mode to a fixed two-armed mode (Fig.~\ref{fig:patterns}b) is straightforward:
When $T$ is decreased below a certain threshold value, bubble $-1$, after collision with bubble $-2$, does not have enough time to move far away from the center before bubble 0 rises to the free surface.
Consequently, bubble 0 collides with bubble $-1$ and bounces towards the opposite direction to the latter's position vector.
Similarly, bubble 1 is emitted in the opposite direction to that of bubble 0, following the path of bubble $-1$.
In this way, the bubbles alternatively join two branches with a constant divergence angle of $\varphi = 180^\circ$, as depicted in Fig.~\ref{fig:patterns}(f).
By conducting simulations with gradually decreased bubble release period, for the case shown in Fig.~\ref{fig:patterns}, this transition is found to occur at $T \approx 80\,\mathrm{ms}$.

\subsection{Multi-armed modes} \label{subsec:multiArmedMode}

\begin{figure}
	\centering
    \includegraphics[width=.9\linewidth]{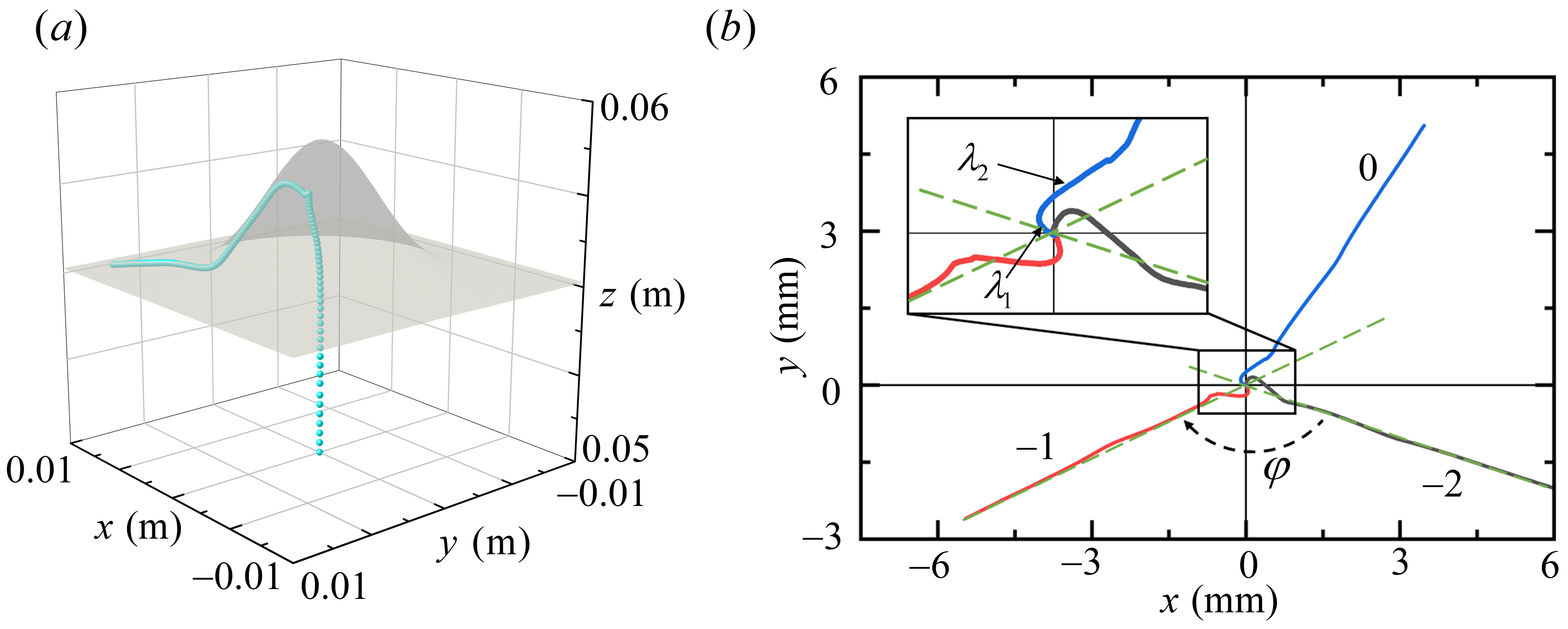}
    \caption{(a) Free surface (gray) and trajectory (cyan) of a single bubble in a five-armed case. (b) Trajectories of three consecutive bubbles projected onto the horizontal plane. The green dashed lines indicate the emission directions of bubbles $-2$ and $-1$.
    The inset is an amplified view of the trajectories in the vicinity of the bubble emergence site.
    }
	\label{fig:trajectory-overview}
\end{figure}

The multi-armed modes occur when the bubble release period is further decreased.
They exhibit richer phenomena compared with the disordered and fixed two-armed situations.
Figures~\ref{fig:patterns}(c) and \ref{fig:patterns}(d) show a fixed five-armed pattern ($T=42$~ms) and a spiral three-armed pattern ($T=36$~ms), respectively.
A 3D view of a single bubble trajectory in the five-armed case
is displayed in Fig.~\ref{fig:trajectory-overview}(a).
The projection of this trajectory onto the horizontal plane is shown in Fig.~\ref{fig:trajectory-overview}(b) alongside the trajectories of two consecutive bubbles. 
An amplified view is provided in the inset, highlighting the curved trajectories of the bubbles near the emergence site.

For patterns with fixed arms, the number $n$ of arms is related to the divergence angle $\varphi$ by 
\begin{equation}
n \varphi = 2 \pi m,
\label{eq:nPhiRelation}
\end{equation}
where $m$ is a small integer.
Without loss of generality, we assume $\varphi$ is positive in Eq.~\eqref{eq:nPhiRelation}.
Eq.~\eqref{eq:nPhiRelation} states that the emissions of $n$ successive bubbles sweep through exactly $m$ complete circles.
Evidently, consecutive bubbles do not necessarily go to adjacent arms.
If Eq.~\eqref{eq:nPhiRelation} is only approximately satisfied, spiral patterns are observed.
For example, with the averaged divergence angles shown in Figs.~\ref{fig:patterns}(g) and \ref{fig:patterns}(h), the fixed five-armed mode corresponds to $5 \times 143.9^\circ \approx 360^\circ \times 2$, whereas the spiral three-armed mode corresponds to $3 \times 124.6^\circ \gtrsim 360^\circ \times 1$.

\begin{figure}
	\centering
    \includegraphics[width=1.\linewidth]{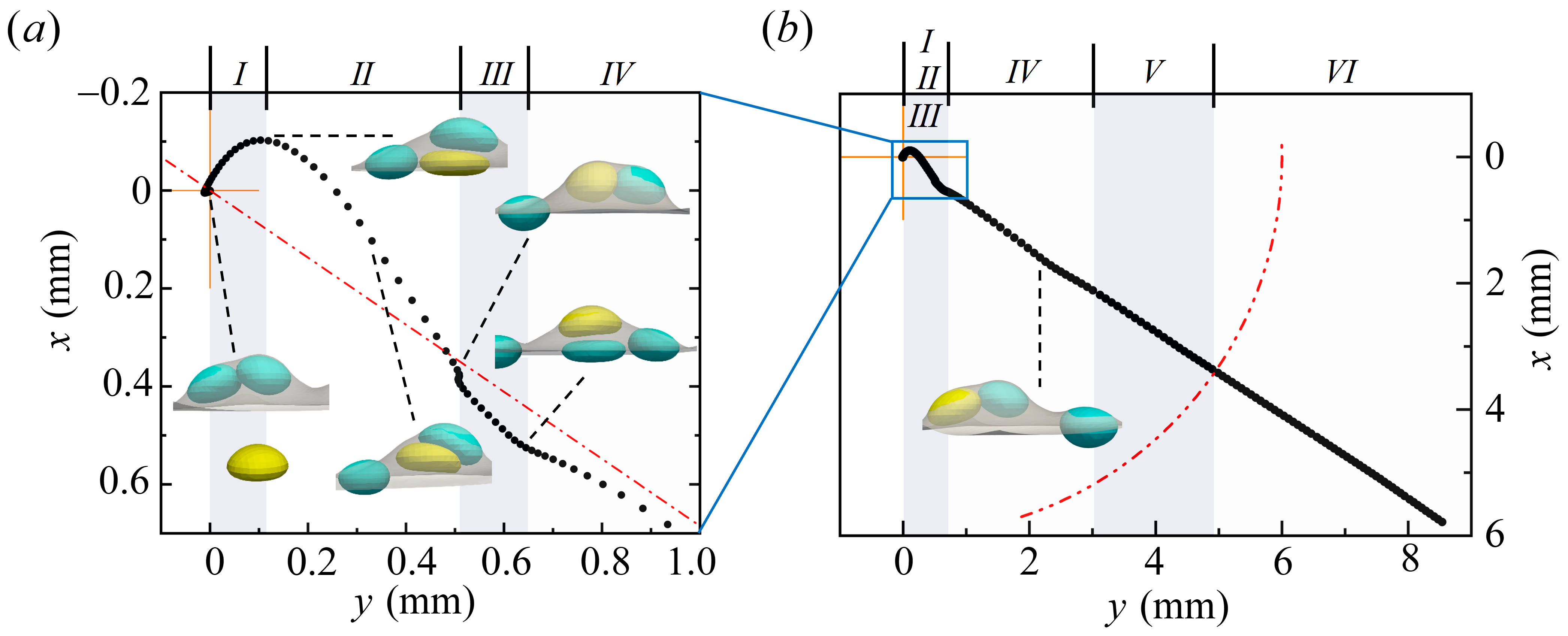}
    \caption{Bubble 0's trajectory projected onto the horizontal plane. Black dots represent positions sampled with equal time intervals. The trajectory is divided into six stages, labeled on the top. 
    Insets show the instantaneous position and shape of bubbles near the free surface (bubble 0 colored yellow). The dash-dotted line in panel (a) indicates the final emission direction of bubble 0. The arc in panel (b) marks the edge of the bump.}
	\label{fig:trajectory}
\end{figure}

The twisted trajectories in Fig.~\ref{fig:trajectory-overview} indicate complex bubble dynamics in the vicinity of the bubble emergence site.
\citet{Yoshikawa2010} note that, despite the arms may rotate, ``the motion of each bubble is purely radial.''
This is true except for in the short-period initial phase of the bubble motion, as can be seen in Fig.~\ref{fig:trajectory-overview}(b).
To elucidate the entire process governing the divergence angle, we divide the trajectory of a single bubble into six distinct stages, labeled I to VI,
as indicated above the trajectory plot in Fig.~\ref{fig:trajectory}. 
This plot displays the bubble's path projected onto the horizontal plane, with the black dots marking positions at equal time intervals (the sparseness of the dots is thus an indicator of the bubble's horizontal-velocity magnitude).
The accompanying insets provide close-up views of the instantaneous configuration of the bubbles near the free surface.
The dynamics of the bubbles and their connection with the processes listed in section~\ref{subsec:overview} are described in detail below.

\begin{figure}
	\centering
    \includegraphics[width=.75\linewidth]{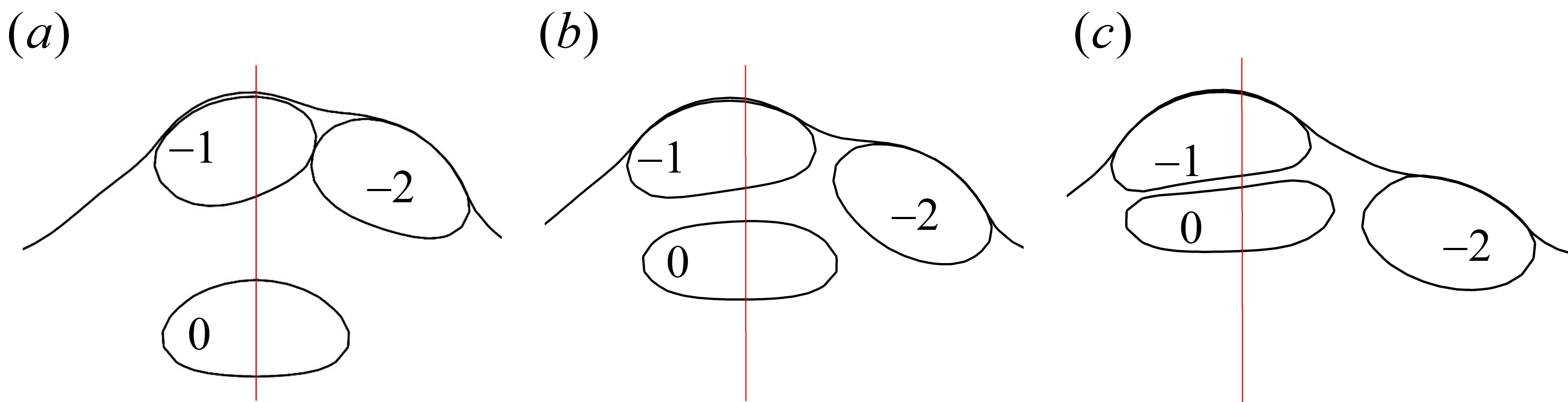}
    \caption{Snapshots of the non-contact repulsion process between bubbles 0 and $-2$, taken in the cross section where the centers of the two bubbles are located. The red line shows the position of the central axis.}
	\label{fig:non-contact}
\end{figure}

Firstly, a significant difference between multi-armed modes and the simpler fixed two-armed mode is that the former cannot be explained solely by binary bubble collisions.
This adds complexity to the analysis of the underlying physics, necessitating the inclusion of process (ii), i.e., non-contact repulsion.
This process is illustrated in Fig.~\ref{fig:non-contact}, where three snapshots of the three-armed situation are shown in chronological order.
During this stage, bubble 0 is still rising below the free surface without colliding with preceding bubbles.
Floating on the free surface with a short distance from the center where the bubbles emerge, bubble $-2$ acts as an obstacle that partially blocks the rightward motion of the liquid flow.
As a consequence of this asymmetry, bubble 0's rising path is deflected to the left, as if it experiences repulsion from bubble $-2$.
Since there is no direct contact between bubbles 0 and $-2$ throughout the interaction, the horizontal displacement of bubble 0 from the central axis is mild.
We denote this displacement by $\bm \lambda_1$.
This process corresponds to stage I in Fig.~\ref{fig:trajectory}(a).
Similar interactions also occur between bubbles 0 and $-1$.
However, since bubble $-1$ is only slightly deviated from the center, its influence on bubble 0's horizontal motion is much weaker than the influence of bubble $-2$.
The repulsion from other floating bubbles is negligible due to their larger distance from bubble 0.
Therefore, $\bm \lambda_1$ is roughly in the opposite direction of bubble $-2$'s position vector in the horizontal plane, as shown in the inset of Fig.~\ref{fig:trajectory-overview}(b).

\begin{figure}
	\centering
    \includegraphics[width=\linewidth]{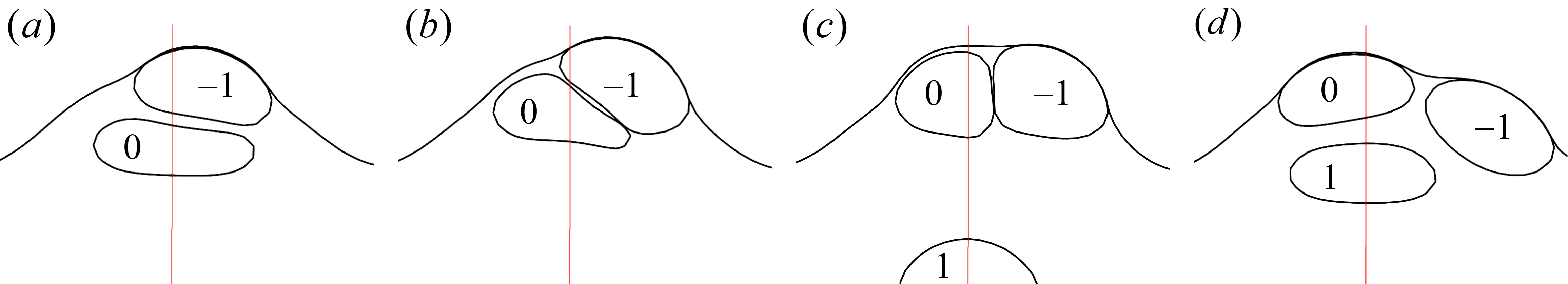}
    \caption{Snapshots of the pair collision process between bubbles 0 and $-1$, taken in the cross section where the centers of the two bubbles are located. The red line shows the position of the central axis.} 
	\label{fig:collision}
\end{figure}

The motion of bubble 0 in stage II is governed by process (i), i.e., pair collision.
Figure~\ref{fig:collision} depicts the collision process between bubbles 0 and $-1$ in the cross-section where the centers of the two bubbles are located.
The snapshot in Fig.~\ref{fig:collision}(a) is taken at about the same time as that in Fig.~\ref{fig:non-contact}(c).
When approaching the free surface, bubble 0 is impeded by bubble $-1$, squeezed and pushed away by it.
During this process, bubble 0 acquires a horizontal displacement roughly in the opposite direction of bubble $-1$'s position vector, which can also be observed in the inset of Fig.~\ref{fig:trajectory-overview}(b).
We denote this displacement by $\bm \lambda_2$.

\begin{figure}
	\centering
    \includegraphics[width=0.65\linewidth]{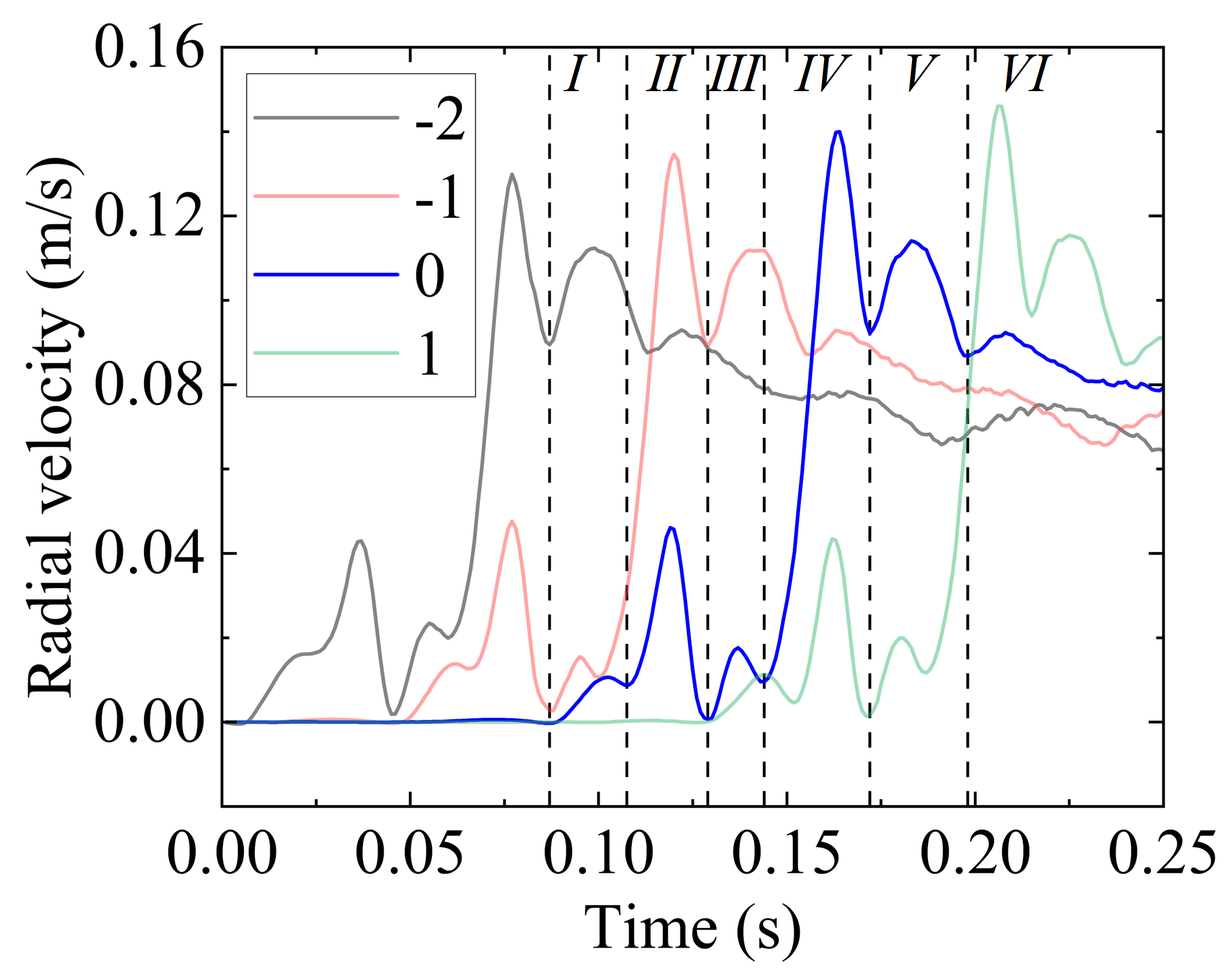}
    \caption{Radial velocity versus time for bubbles $-2$, $-1$, 0, and 1 (from left to right). Stages are labeled with respect to bubble 0.}
	\label{fig:velocity1}
\end{figure}

Although processes (i) and (ii) both have effects to accelerate bubble 0, the bubble's radial velocity falls back to nearly zero by the end of stage II.
This can be seen from Fig.~\ref{fig:velocity1}, where the radial velocities of bubbles $-2$, $-1$, 0, and 1 are shown as functions of time.
This deceleration of bubble 0 is due to process (iii) mentioned in section~\ref{subsec:overview}, i.e., the confinement of the bump.
As revealed in Fig.~\ref{fig:collision}(c), after collision with bubble $-1$, bubble 0 is located in the very top region of the bump.
Thus, its horizontal motion is restricted by the capillary effect of the deformed and more vertical side surface of the bump.

It is noteworthy that, up to this point, bubble 0's displacement from the central axis is no greater than half its radius.
However, the final emission direction (the red dash-dotted line in Fig.~\ref{fig:trajectory}a) of the bubble has been essentially determined.
All the remaining stages play a role to ``launch'' this nearly stationary bubble and transport it in the same direction as its current position vector.

The collision with bubble 0 also promotes bubble $-1$ to leave the central region.
As a consequence, bubble 0 has more room to flatten, and the bump surface becomes less steep.
This relieves the constraint imposed by the bump, allowing bubble 0 to regain a small velocity away from the center advected by the radial flow mentioned earlier. 
This process corresponds to stage III in Figs.~\ref{fig:trajectory} and \ref{fig:velocity1}.

Stage IV is characterized by the pair collision between bubbles 0 and 1, which continues to accelerate bubble 0.
The radial velocity of bubble 0 rapidly reaches its maximum value over the entire bubble motion, followed by a short period of deceleration caused again by the confinement of the deformed bump.
The instantaneous configuration is similar to that shown in Fig.~\ref{fig:collision}(c), but with bubble numbers $-1$ and 0 replaced by 0 and 1, respectively.

\begin{figure}
	\centering
    \includegraphics[width=.5\linewidth]{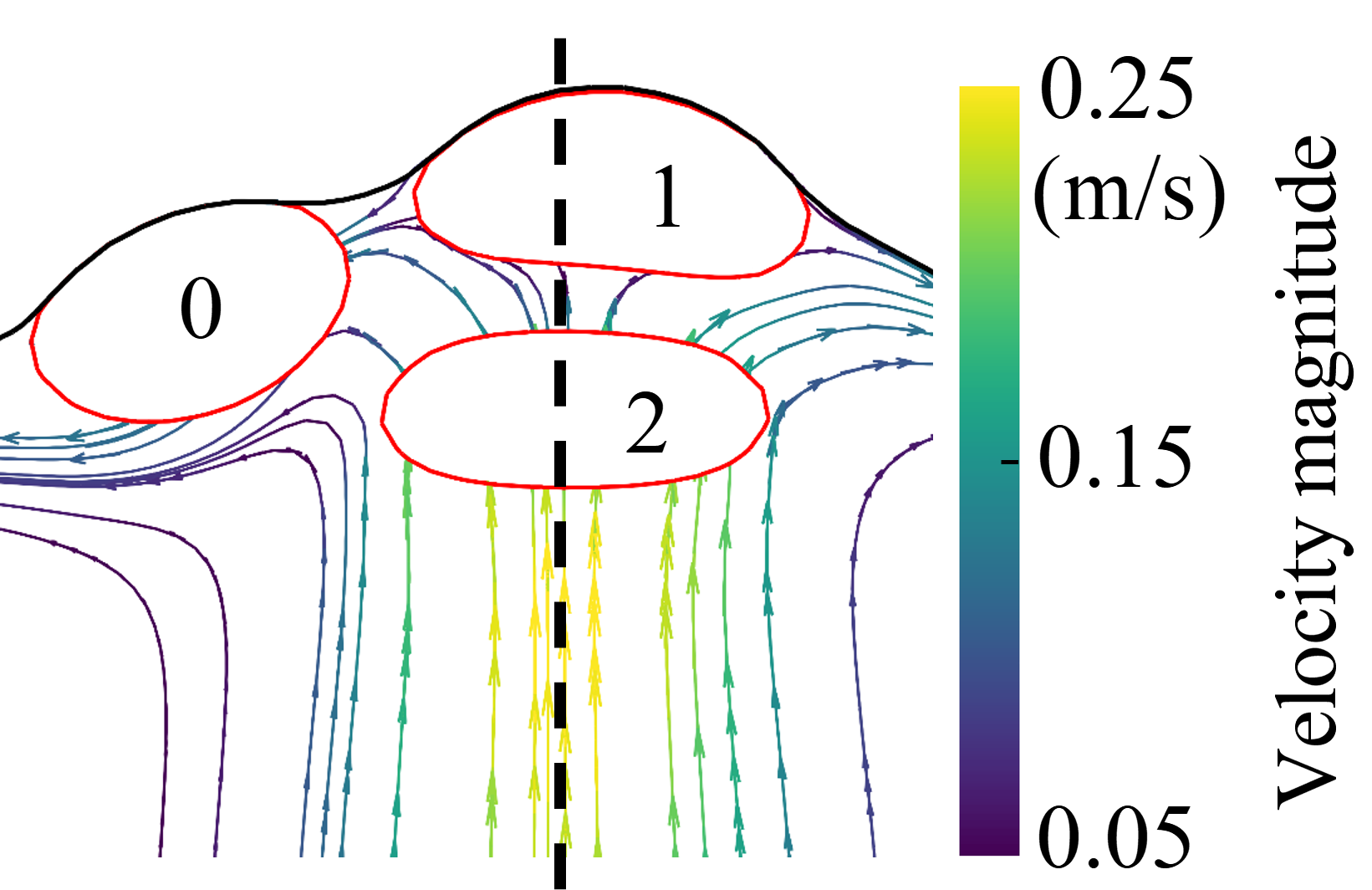}
    \caption{Snapshot of a cross-sectional view of the bubbles near the bump,
    with streamlines colored according to the velocity magnitude.
    The free surface is represented by the black solid line.
    The dashed line indicates the central axis.}
	\label{fig:flowField1}
\end{figure}

\begin{figure}
	\centering
    \includegraphics[width=0.65\linewidth]{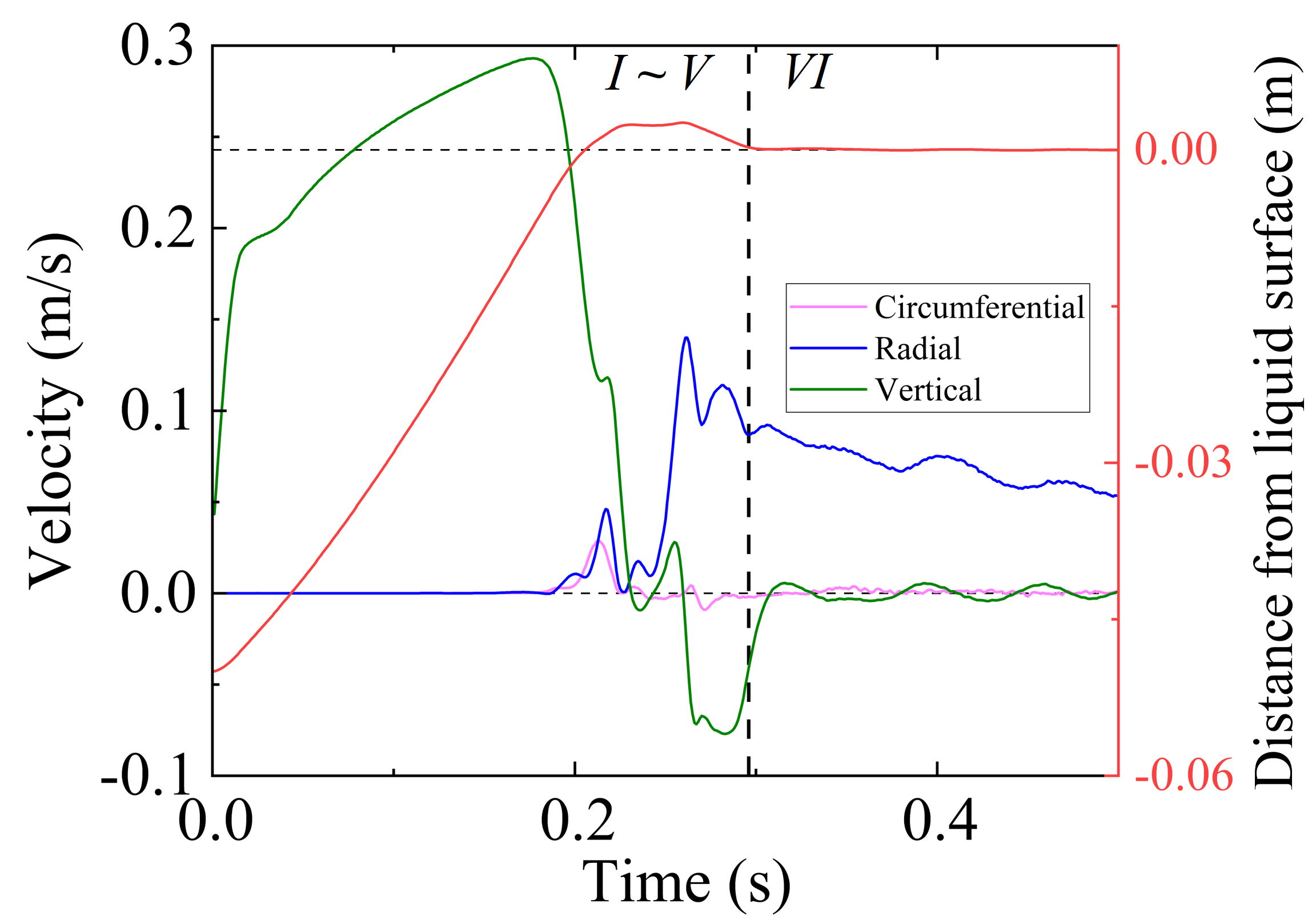}
    \caption{Time dependence of bubble 0's radial (blue), vertical (green), and circumferential (pink, clockwise positive) velocities.
    The red line represents the distance between the bubble mass center and the undisturbed free surface.
    Time is set to zero at the instant of bubble release.}
	\label{fig:velocity2}
\end{figure}

Being pushed away from the center by bubble 1, the subsequent motion of bubble 0 in stage V is driven by the divergent radial flow (process (iv) in section~\ref{subsec:overview}).
This effect of flow advection is demonstrated by the streamlines shown in Fig.~\ref{fig:flowField1}.
The bubble continues moving within the bump until it encounters the edge of it, where the bubble experiences a temporary deceleration due to the transition between the bumped and horizontal free surface.
This can be seen in Fig.~\ref{fig:velocity2}, where the vertical dashed line indicates that the local minimum radial velocity occurs exactly when the bubble reaches the flat liquid surface from a higher elevation.

\begin{figure}
	\centering
    \includegraphics[width=.5\linewidth]{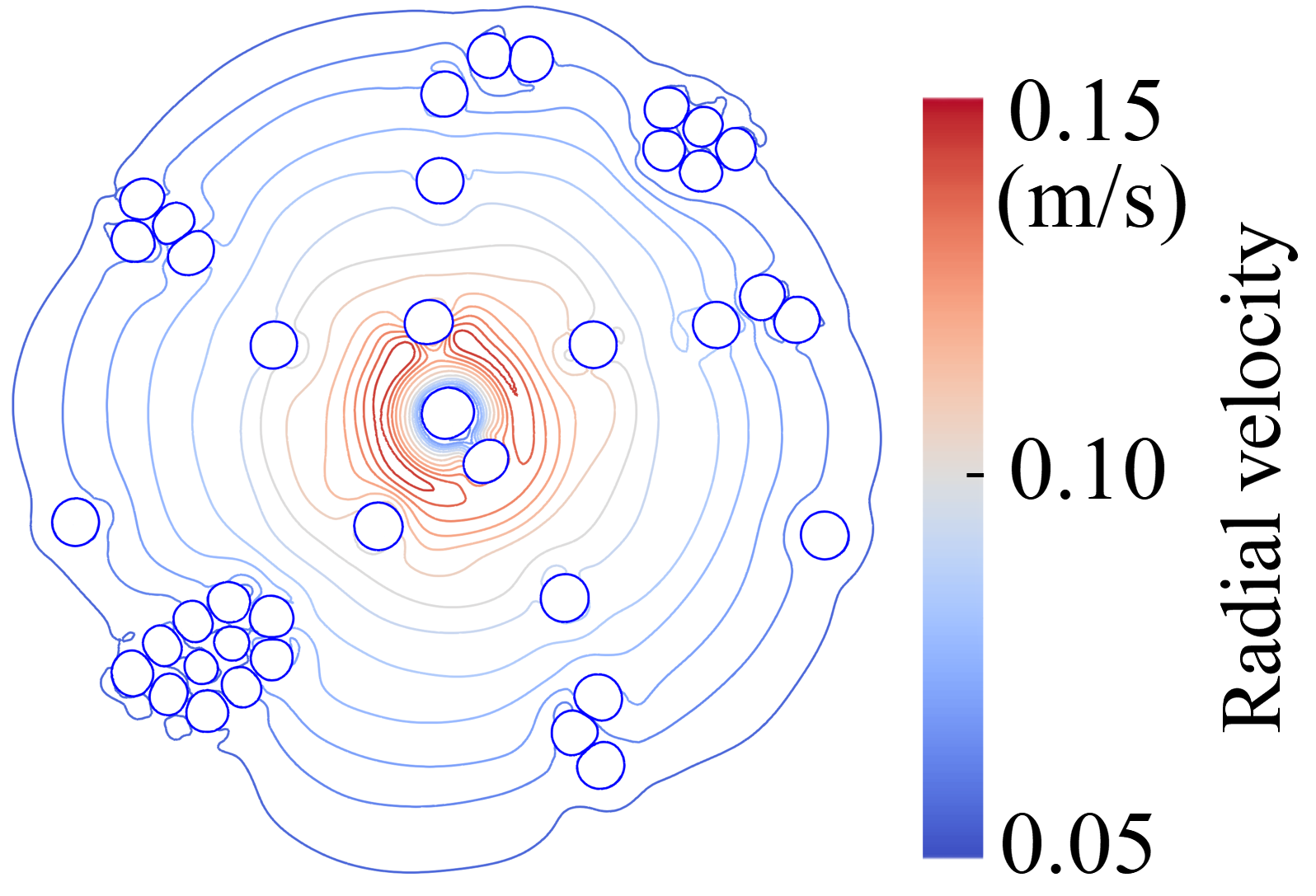}
    \caption{Contour lines of the liquid's radial velocity in a horizontal plane slightly below the free surface. The small circles are bubbles floating on the surface.}
	\label{fig:flowField2}
\end{figure}

After escaping from the bump, in stage VI, bubble 0 assumes a purely radial motion.
Figure~\ref{fig:flowField2} presents the contour lines of the liquid's radial velocity in a horizontal plane slightly below the free surface.
According to this figure, the radial velocity of the bubbles is smaller than that of the surrounding liquid, revealing that the bubbles are passively advected by the divergent liquid flow in the surface layer.
During this stage, both the radial and vertical velocity components of the bubble undergo mild fluctuations over time, as can be observed in Fig.~\ref{fig:velocity2}.
These fluctuations are likely caused by the outward-propagating annular surface waves faintly visible in the top row of Fig.~\ref{fig:patterns}.
Due to volume conservation and viscous dissipation, the radial flow weakens with increasing radius.
The bubbles gradually decelerate and eventually stop at some distance from the center,
forming discrete clusters, as is evident in Figs.~\ref{fig:experiment}, \ref{fig:patterns}(c), and \ref{fig:patterns}(d).

\section{A heuristic model}\label{sec:theory}
\begin{figure}
	\centering
    \includegraphics[width=\linewidth]{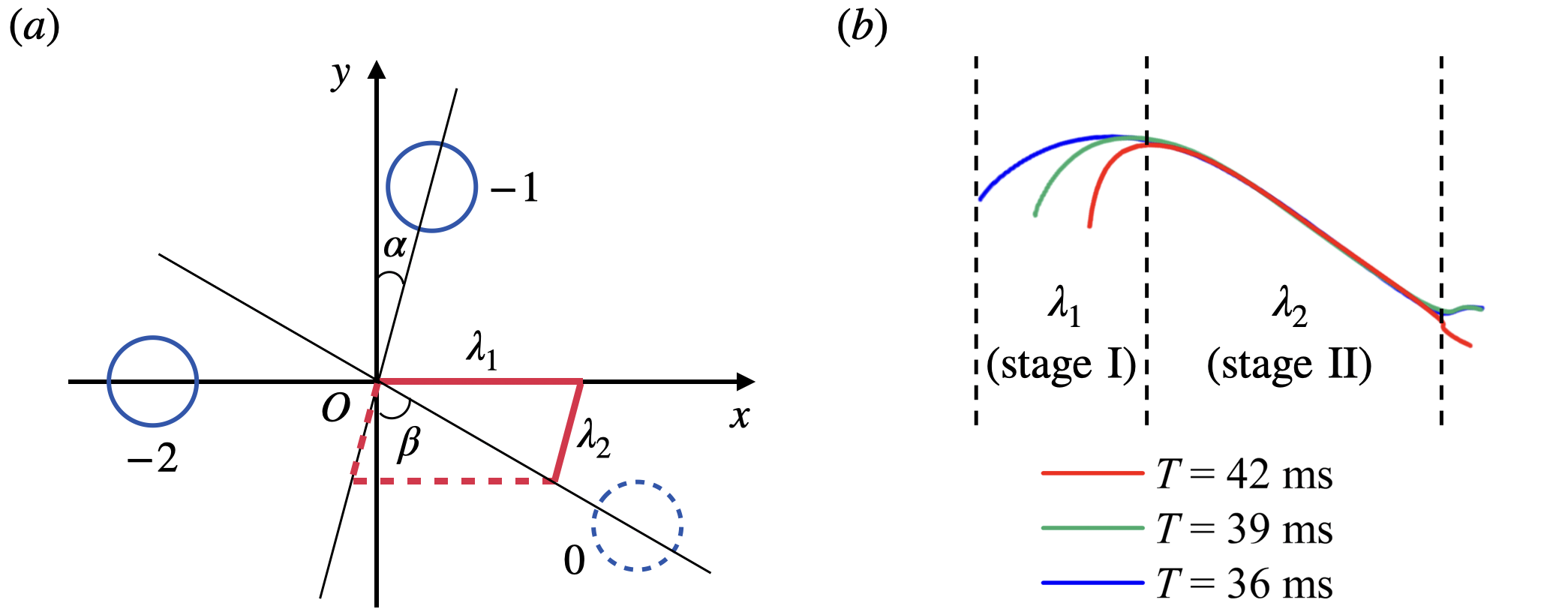}
    \caption{(a) Geometric configuration of bubbles on the free surface (bubble size and position for illustrative purposes only).
    (b) Simulated bubble trajectories projected onto the horizontal plane after appropriate translation and rotation. 
    }
	\label{fig:theory}
\end{figure}

We propose a simple model to describe the dependence of the divergence angle $\varphi$ on the bubble release period $T$, provided that all other parameters are fixed.
Without loss of generality, we assume the motion of bubble $-2$ is in the direction of $-x$, as illustrated in Fig.~\ref{fig:theory}(a).
Based on the discussions in section~\ref{subsec:multiArmedMode}, in stages I and II, the motion of bubble 0 is mainly influenced by bubbles $-2$ and $-1$, with the displacements occurring in the opposite directions of these two bubbles.
Therefore, $\bm \lambda_1$ and $\bm \lambda_2$ are plotted in Fig.~\ref{fig:theory}(a) parallel to the reverse extensions of the two bubbles' position vectors.
In later stages, bubble 0 is launched in the direction of the vector sum of $\bm \lambda_1$ and $\bm \lambda_2$.
In order that the successive bubbles form a regular pattern, the angles between bubbles 0 and $-1$ and between bubbles $-1$ and $-2$ are both equal to $\varphi$.
With introduction of the variable $\alpha = \varphi - \pi/2$, one obtains $\beta = \pi/2 - 2\alpha$, where $\beta$ is defined in Fig.~\ref{fig:theory}(a).
Denoting the magnitude of $\bm \lambda_1$ and $\bm \lambda_2$ by $\lambda_1$ and $\lambda_2$, respectively, the geometric constraint requires $\tan \beta = (\lambda_1 - \lambda_2 \sin\alpha)/(\lambda_2 \cos\alpha)$.
Elimination of the auxiliary variables $\beta$ and $\alpha$ yields
\begin{equation}
\varphi = \frac{\pi}{2} + \arcsin \frac{\lambda_2}{2 \lambda_1}.
\label{eq:geometric}
\end{equation}
Based on Eq.~\eqref{eq:geometric},
when $\lambda_2/\lambda_1 \rightarrow 2^-$, $\varphi \rightarrow \pi$, corresponding to the two-armed mode limit;
when $\lambda_2/\lambda_1 \rightarrow 0^+$, $\varphi \rightarrow \pi/2$, indicating a four-armed mode.
However, this latter situation is unlikely to occur because of the fact that while $\lambda_2$ is finite in general, $\lambda_1$, caused by non-contact repulsion, cannot be excessively large.
The situation of $\lambda_2/\lambda_1 > 2$ is not covered by Eq.~\eqref{eq:geometric}, either can it be described by the geometrical configuration shown in Fig.~\ref{fig:theory}(a).
Actually, it represents the scenario that the non-contact repulsion between bubbles $-2$ and 0 is too small to influence the pattern formation, and the divergence angle is dominated by the pair collision between bubbles $-1$ and 0.
Thus, the fixed two-armed mode or disorder mode is expected.

It is useful to further relate $\lambda_1$ and $\lambda_2$ to the release period $T$.
The two displacements for three cases with different $T$ are shown in Fig.~\ref{fig:theory}(b) after appropriate translation and rotation within the horizontal plane.
It is noteworthy that, while $\lambda_1$ varies, $\lambda_2$ is nearly the same for all the three cases.
This is because $\lambda_2$ is caused by the pair collision within the bump.
The strong effects of squeezing and bump confinement render $\lambda_2$ dominated by geometric factors such as the bubble and bump sizes.
Therefore, it is reasonable to consider $\lambda_2$ as a constant that is independent of $T$.
On the other hand, $\lambda_1$, which results from non-contact repulsion, is expected to be inversely related to the average distance between bubbles 0 and $-2$ as bubble 0 approaches the free surface.
This distance, in turn, is positively correlated with $T$.
Based on this analysis, a very simple relation between $\lambda_1$ and $T$ is proposed as $\lambda_1 \propto 1/(a T - b)$, where the coefficients $a$ and $b$ may depend on parameters such as the bubble size, liquid viscosity, and surface tension coefficient.
This relation can be substituted into Eq.~\eqref{eq:geometric} while keeping $\lambda_2$ constant.
After including the fixed two-armed mode situation, the model is expressed as
\begin{equation}
\varphi = \begin{cases}
\displaystyle\frac{\pi}{2} + \arcsin \left( a T - b \right),\quad &T \le \displaystyle\frac{b+1}{a},\\
\pi,\quad &T>\displaystyle\frac{b+1}{a}.
\end{cases}
\label{eq:model}
\end{equation}

\begin{figure}
	\centering
    \includegraphics[width=1.\linewidth]{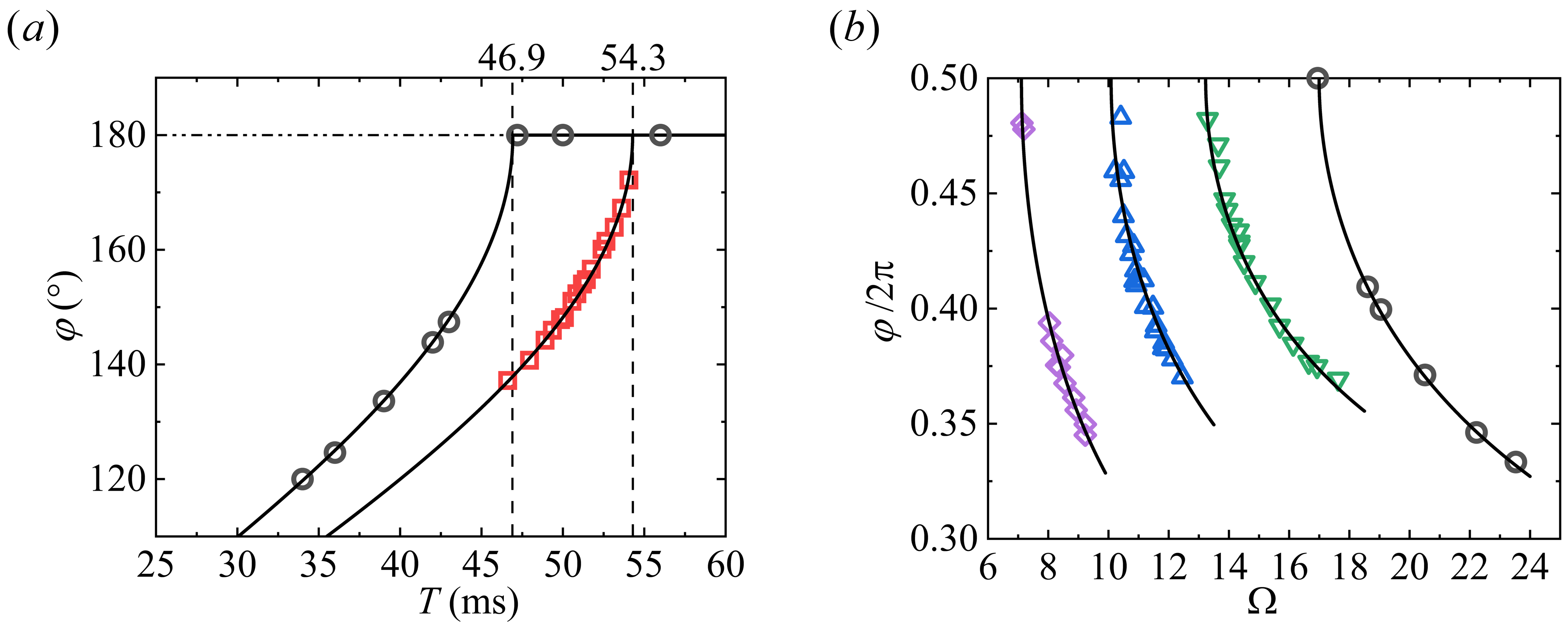}
    \caption{
    (a) Divergence angle $\varphi$ versus release period $T$. 
    Red squares denote experimental data extracted from Fig.~1.2(a) in Ref.~\citep{Yoshikawa2012}, with $\mathit{Oh} \approx 0.099$ and $\mathit{Bo}=17$.
    (b) Divergence angle versus dimensionless frequency $\Omega$.
    The three data sets on the left are experimental results extracted from Fig.~5(a) in Ref.~\citep{Yoshikawa2010}.
    From left to right, the Ohnesorge and Bond numbers, estimated based on the available information in Ref.~\citep{Yoshikawa2010}, are approximately 0.26 and 23, 0.21 and 18, 0.11 and 11, respectively.
    In both panels, the present numerical results are denoted by black circles; solid curves represent predictions using Eq.~\eqref{eq:model} or \eqref{eq:model2}.
    }
	\label{fig:yoshikawa}
\end{figure}

Eq.~\eqref{eq:model} is valid unless $T$ is too large, in which case disordered, instead of regular, patterns, will be observed.
Figure~\ref{fig:yoshikawa}(a) shows the averaged $\varphi$ as a function of $T$ for eight simulation cases (black circles) and the experimental data from Ref.~\citep{Yoshikawa2012} (red squares).
Notably, with $a$ and $b$ fitted for each data set, the predictions of Eq.~\eqref{eq:model} (solid lines) have a remarkable consistency with both the numerical and experimental results.
The fitted coefficients for the two curves, from left to right, are $a=0.039$\,ms$^{-1}$, $b=0.83$, and $a=0.035$\,ms$^{-1}$, $b=0.90$, respectively.

Eq.~\eqref{eq:model} can be rewritten in dimensionless form as:
\begin{equation}
\displaystyle\frac{\varphi}{2\pi} = \begin{cases}
\displaystyle\frac{1}{4} + \displaystyle\frac{1}{2\pi}\arcsin \left( \displaystyle\frac{A}{\Omega} - b \right),\quad &\Omega \ge \displaystyle\frac{A}{b+1},\\
\displaystyle\frac{1}{2},\quad &\Omega<\displaystyle\frac{A}{b+1},
\end{cases}
\label{eq:model2}
\end{equation}
where $A= \rho_l d^2 a/\mu_l$, and $\Omega=\rho_l d^2 / (\mu_l T)$ is the dimensionless bubble release frequency defined in Ref.~\citep{Yoshikawa2010}.
The dependence of $\varphi/(2\pi)$ on $\Omega$ is shown in Fig.~\ref{fig:yoshikawa}(b).
The present numerical results are still denoted by black circles.
The other three data sets correspond to experimental data extracted from Fig.~5(a) in Ref.~\citep{Yoshikawa2010}.
The results of Eq.~\eqref{eq:model2} are shown as black curves with $A$ and $b$ fitted for each set of data.
The values of $A$ and $b$ for the four curves in Fig.~\ref{fig:yoshikawa}(b), from left to right, are: 13.3 and 0.87, 16.5 and 0.64, 17.9 and 0.35, 31.1 and 0.83, respectively.
Good agreement is observed between the model predictions and all the data sets.

\begin{figure}
	\centering
    \includegraphics[width=\linewidth]{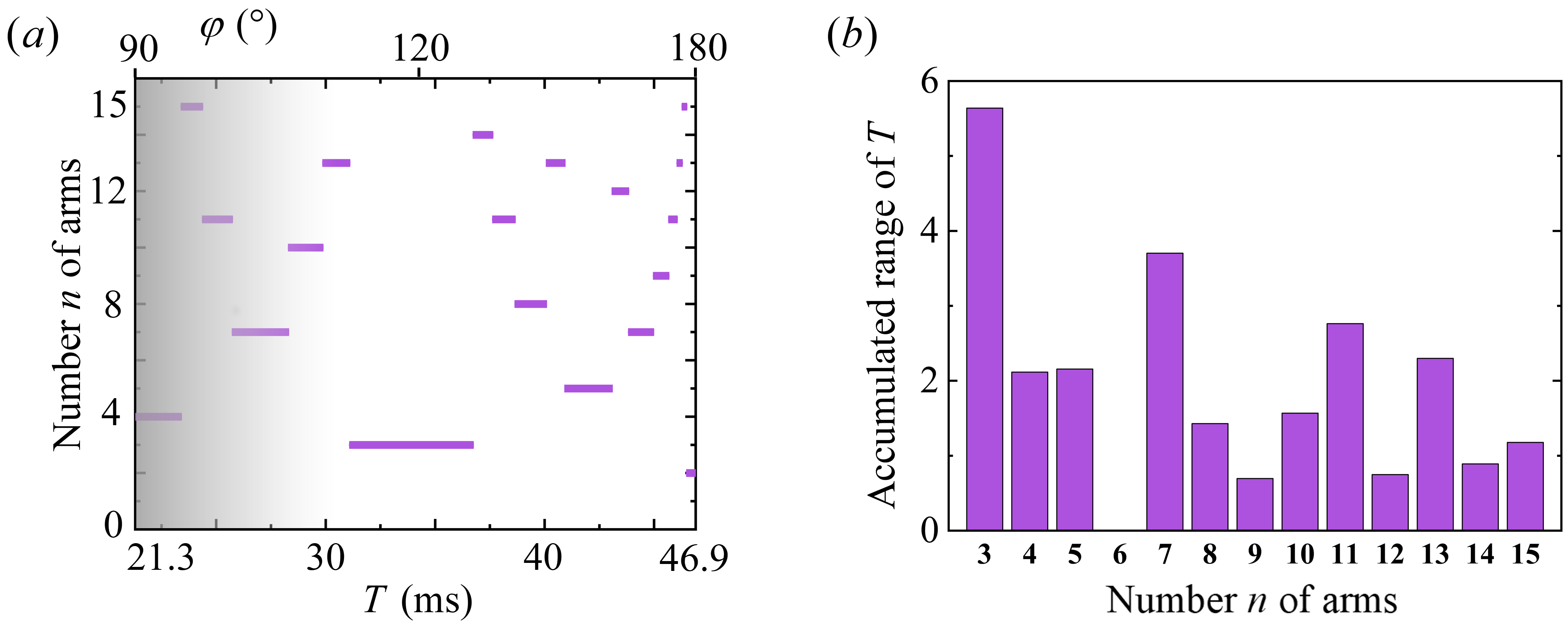}
    \caption{(a) Number of observed arms estimated using Eq.~\eqref{eq:model} (with $a=0.039$\,ms$^{-1}$ and $b=0.83$) and Eq.~\eqref{eq:nPhiRelation} as a function of bubble release period $T$.
    Cases in the gray area are unlikely to occur.  
    (b) Accumulated range of $T$ within which $n$-armed mode can be observed.}
	\label{fig:probablity}
\end{figure}

In the following, we consider the possibility of observing different multi-armed patterns in reality.
Figure~\ref{fig:probablity}(a) shows the number $n$ of arms that is expected to be seen as the release period $T$ varies.
This mapping is obtained by calculating the corresponding $\varphi$ at a given $T$ using Eq.~\eqref{eq:model}, then finding $n$ within the range $2 \le n \le 15$ that best fits Eq.~\eqref{eq:nPhiRelation} with all possible $m$.
The upper bound of 15 is set considering that for very large $n$, the finite-sized bubbles are so crowded that the arms become difficult to distinguish in practice.
It can be seen from Fig.~\ref{fig:probablity}(a) that each $n$ may correspond to several discontinuous intervals of $T$.
The accumulated range of $T$ for a given $n$ is shown in Fig.~\ref{fig:probablity}(b).
The probability of appearances of $n$-armed mode can thus be estimated based on the height of the bars.

According to Fig.~\ref{fig:probablity}, a six-armed mode seldom exists.
In addition, due to the reason discussed earlier, situations near the left edge of Fig.~\ref{fig:probablity}(a) ($\varphi \gtrsim 90^\circ$ with $\lambda_2/\lambda_1 \rightarrow 0^+$) cease to be possible, reducing the possibility of $n=4$, 7, 11, etc.
The above analyses align well with our experience that the three- and five-armed patterns are easier to achieve than other multi-armed modes by adjusting the period $T$ in the numerical simulations.

\section{Summary and conclusions} \label{sec:conclusion}

This work elucidates the formation 
of regular surface patterns induced by chains of small rising bubbles.
Direct numerical simulations reveal how distinct morphologies stem from non-contact bubble repulsion, pair collisions, free-surface deformation, and ambient flow advection.
We establish a heuristic model relating the number of pattern arms (represented by the divergence angle $\varphi$) to the bubble release period $T$, showing quantitative agreement with both simulation data and experimental results from the literature.
While this heuristic model is not derived from
first-principle fluid dynamics, it bridges observed geometric patterns with causal hydrodynamic processes,
offering a predictive tool for the self-organized pattern formation in non-coalescing bubble systems.
The main predictions of the model include:
(1) the existence of stable, discrete surface patterns with fixed arm numbers;
(2) a quantitative relationship between the divergence angle of successive bubbles (thus the arm number) and their release period;  
and (3) preferential occurrence of specific patterns, with three- and five-armed cases being the most probable.
It is worthwhile to mention that, the focus of this study is the general relationship between $\varphi$ and $T$. 
Effects of other parameters--such as bubble diameter, liquid viscosity, and surface tension coefficient--are implicitly included in the coefficients $a$ and $b$ of the proposed model (Eq.~\eqref{eq:model}). 
If bubbles have not reached terminal velocity near the free surface, liquid depth becomes an additional parameter that has a quantitative influence on $a$ and $b$; but all essential processes
responsible for pattern formation remain unchanged.

A key feature of the regular patterns is the lateral dispersion--instead of aggregation--of surface bubbles.
This suggests potential applications in interfacial heat transfer enhancement via foam suppression and contaminant removal through spiral-pattern-driven surface sweeping.
Moreover, the patterns' sensitivity to operational parameters enables real-time monitoring of gas flow rates, bubble release intervals, and fluid properties.
As a minimal model of self-organization, this study demonstrates how localized dynamics govern emergent macroscopic order, offering insights for understanding, designing, and controlling collective behaviors in broader self-organized systems.

\vspace{1em}
\noindent \textbf{ACKNOWLEDGEMENTS}

This work was supported by the National Natural Science Foundation of China (Grant No. 12202441) and the Xiaomi Young Scholar Program.
D. Li appreciates the advice from Dr. Petr Karnakov on the use of Aphros.
The authors also thank Dr. Zibo Ren for his participation in the early stage of this study.

\appendix*
\section{Mesh-Independence Assessment} \label{sec:appendix}

\begin{figure}[h]
	\centering
    \includegraphics[width=\linewidth]{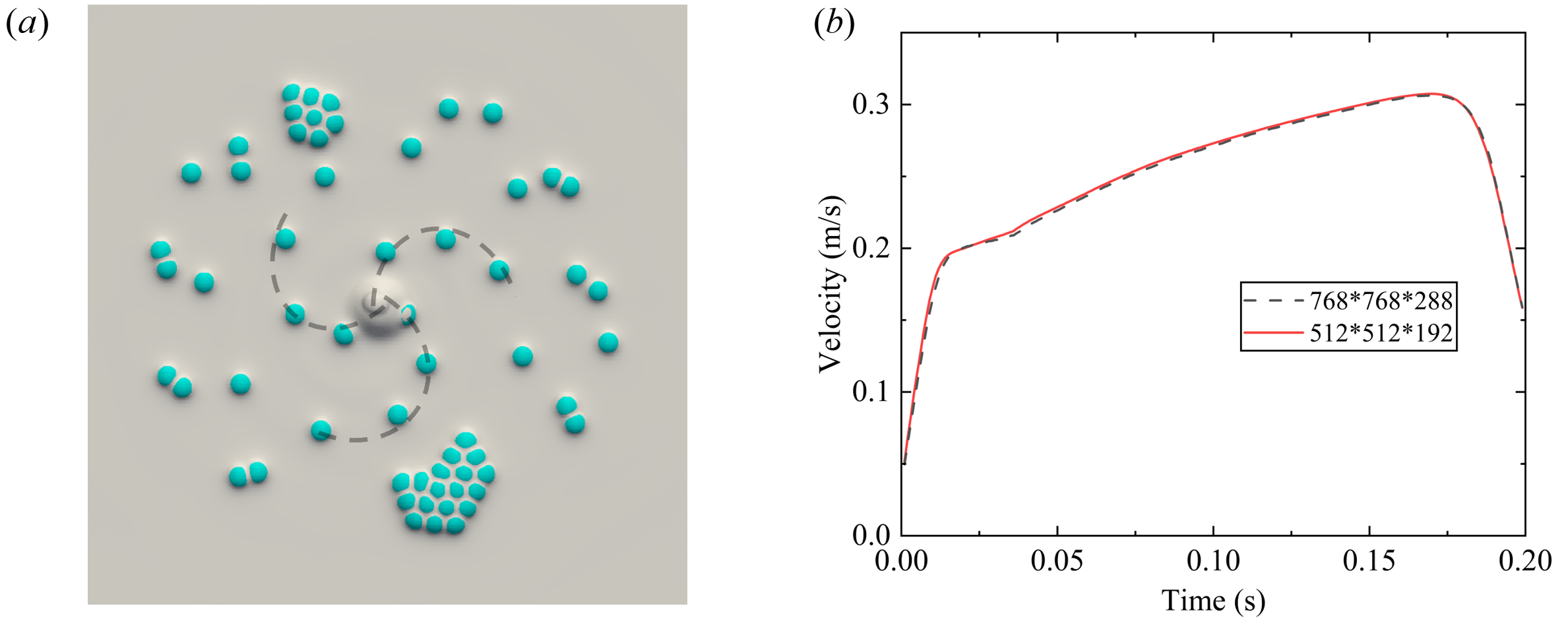}
    \caption{(a) Surface pattern obtained with a mesh resolution of $768\times768\times288$ for the three-armed case shown in Fig.~\ref{fig:patterns}.
    (b) Temporal evolution of the bubble rise velocity. The fine- and coarse-mesh results are denoted by dashed and solid lines, respectively.}
	\label{fig:Grid_Convergence}
\end{figure}

The spiral three-armed pattern in Fig.~\ref{fig:patterns}(d) was obtained with a base mesh resolution of $512\times512\times192$.
To validate this result, an additional simulation for the same case is conducted using a finer mesh ($768\times768\times288$).
This higher-resolution computation requires more than 20 days if running in parallel on 128 CPU cores.
The resulting pattern is presented in Fig.~\ref{fig:Grid_Convergence}(a).
Although the pattern rotates in the opposite direction (the direction has no preference, as it is determined by random disturbances during initial bubble interactions), the average divergence angles agree closely (128.8$^\circ$ versus $124.6^\circ$), with a difference of less than 3.5\%.

Moreover, Fig.~\ref{fig:Grid_Convergence}(b) compares the temporal evolution of the rise velocity for a specific bubble in the bubble chain.
The near overlap of the curves from both meshes further justifies the use of the base mesh resolution.

\bibliography{apstemplate}

\end{document}